\UseRawInputEncoding

\documentclass[aps,prb,twocolumn,groupedaddress,showpacs, citeautoscript,superscriptaddress]{revtex4-1}
\usepackage{graphicx}
\usepackage{dcolumn}
\usepackage{bm}
\usepackage{amsmath,amsfonts,amssymb,amsthm,verbatim}
\usepackage[ansinew]{inputenc}
\usepackage{color}

\usepackage{epsfig}

\begin{document}

\title{Symmetry-enforced nodal lines in the band structures of vacancy-engineered graphene}


\author{Matheus S. M. de Sousa}

\affiliation{Department of Physics, PUC-Rio, 22451-900 Rio de Janeiro, Brazil}

\author{Fujun Liu}

\affiliation{Nanophotonics and Biophotonics Key Laboratory of Jilin Province, School of Physics, Changchun University of Science and Technology, Changchun, 130022, P.R.China}

\author{Mariana Malard}

\affiliation{Faculdade UnB Planaltina, Universidade de Bras\'{\i}lia, Bras\'{\i}lia-DF, Brazil}

\author{Fanyao Qu}

\affiliation{Instituto de F\'{i}sica, Universidade de Bras\'{i}lia, Bras\'{i}lia-DF, Brazil}

\author{Wei Chen}

\affiliation{Department of Physics, PUC-Rio, 22451-900 Rio de Janeiro, Brazil}

\date{\today}

\begin{abstract}

We elaborate that single-layer graphene with periodic vacancies can have a band structure containing nodal lines or nodal loops, opening the possibility of graphene-based electronic or spintronic devices with novel functionalities. The principle is that by removing carbon atoms such that the lattice becomes nonsymmorphic, every two sublattices in the unit cell will map to each other under glide plane operation. This mapping yields degenerate eigenvalues for the glide plane operation, which guarantees that the energy bands must stick together pairwise at a boundary of the Brillouin zone. Moving away from the Brillouin zone boundary causes the symmetry-enforced nodal lines to split, resulting in accidental nodal lines caused by the crossings of the split bands. Moreover, the density of states at the Fermi level may be dramatically enhanced if the nodal lines crosses the Fermi level. The nodal lines occur a variety of vacancy configurations even in the presence of Rashba spin-orbit coupling. Finally, our theory also explains the nodal loops surrounding the entire Brillouin zone of a chevron-type nanoporous graphene fabricated in a recent experiment.


\end{abstract}

\maketitle

\section{Introduction}

A nodal-line semimetal (NLSM) is a novel phase of matter, characterized by band crossings along lines, loops, or even circles in the Brillouin zone (BZ)\cite{Fang2015}. Both theoretical and experimental results have demonstrated that NLSMs possess various interesting properties, such as chiral anomaly \cite{Zyuzin2012}, extremely large magnetoresistance \cite{Gao2017}, photo-induced anomalous Hall effects \cite{Liu2018},  high thermal conductivity, giant intrinsic charge mobility, non-Abelian statistics and superconductivity \cite{Alicea2011}, which have strongly motivated the research on NLSMs\cite{Feng2017,Liu2018,Nie2020,Zunge2018}. Concerning the mechanisms for the formation of these nodal lines and nodal loops, they can be either accidental or symmetry-enforced\cite{Zuo2019,Schoop2016,Fang2015,YangSY2018}. The former are related to various spatial or nonspatial symmetries, such as the Dirac nodal lines protected by reflection, space-time inversion, or rotation symmetry, and can be adiabatically destroyed by tuning parameters of the material, such as spin-orbit coupling (SOC). Symmetry enforced NLSM phases \cite{Schoop2016,Fang2015}, on the other hand, emerge in crystals with nonsymmorphic symmetries, which warrants global stability to the associated nodal lines.

A recent work proves a remarkable principle to engineer nodal lines in two-dimensional (2D) materials: By periodically removing atoms such that the lattice becomes nonsymmorphic, the material becomes a robust NLSM\cite{Liu21,Liu21_2}. As a case study, it was shown that the band structures of a number of vacancy-engineered borophenes display nodal lines originated from a nonsymmorphic glide-plane symmetry. In this paper, we demonstrate that vacancy engineering can in fact be applied to another important 2D material, namely the single-layer graphene. As the first member of the 2D family, graphene exhibits a variety of extraordinary properties with huge impact to applied research, including unprecedented high strength and flexibility, ultralow weight, ultrahigh carrier nobilities, high optical transparency and high thermal conductivity \cite{Castro-2009}. However, these physical properties rely on the linear Dirac cones at low energy which are rather frail to SOC, especially the Rashba SOC, which may hinder its application in SOC-based devices. In particular, Rashba SOC is known to be tunable by a gate voltage in graphene/transition-metal dichalcogenide heterostructures\cite{Yang16,Wang16,Yang17,Offidani17,Safeer19,Ghiasi19,Benitez20}, which may be used to engineer a variety of spintronic effects, such as the recently discovered edge current, edge spin current, and bias-voltage free spin torque in geometrically confined graphene\cite{deSousa21}. A vacancy-engineered graphene that supports nodal lines even in the presence of Rashba SOC could thus open up new possible functionalities. Moreover, we find that the density of states (DOS) near the Fermi level is dramatically enlarged if the nodal lines passes the Fermi level, which is expected to strongly impact the electronic and magnetic properties of the proposed structures. This connects with studies of twisted bilayer graphene which hosts superconductivity when tuned to special ``magic angles" at which isolated and relatively flat bands appear \cite{Cao-2018}.


Besides nodal lines, here we demonstrate that vacancy-engineering also allows to generate nodal loops surrounding the entire boundary of the BZ. Interestingly, the scheme additionally gives rise to accidental nodal lines and loops inside the BZ that are robust to Rashba SOC.
Regarding the feasibility of our proposal, various experimental techniques such as self-aligned anisotropic etching\cite{Shi11}, copolymer lithography\cite{Bai10}, nano-network masking\cite{Jung14}, nanosphere lithography\cite{Wang13}, and nitrogenation\cite{Mahmood-2018} have been employed to fabricate graphene with vacancies, often called graphene nanomesh or holey graphene. In particular, we will elaborate that the nodal loops in fact have already been realized in a recent experiment that fabricates a nonsymmorphic chevron-type nanoporous graphene\cite{Jacobse20}, although this feature seems to be overlooked. Additionally, to further illustrate the generality of our scheme, we show that a vacancy-engineered square lattice also supports nodal loops. Thus we anticipate that the proposed vacancy engineering principle may be further exploited to design Rashba SOC-active nodal-line spintronic or electronic devices in a wide variety of 2D materials.

Naturally nonsymmorphic materials (i.e. that contain glide-plane or screw-axis symmetries without lattice engineering) are expected to possess symmetry-enforced band degeneracies\cite{Young15,Yamakage16,Zhao16,Wieder18}. Three-dimensional realizations of such materials have been predicted in hexagonal compounds\cite{Zhang18}. Our aim is to put forward a simple and practical method to create 2D nonsymmorphic materials which contains multiple nodal-lines or nodal-loops. We argue theoretically and illustrate numerically that it can done simply by periodically removing atoms from common monoatomic and nonmagnetic sheets, using graphene as a concrete example.


\section{NLSM phases in vacancy-engineered graphene}

\subsection{Vacancy-engineered graphene with a single glide plane \label{sec:single_glide}}

\begin{figure}[ht]
\includegraphics[width=0.99\columnwidth]{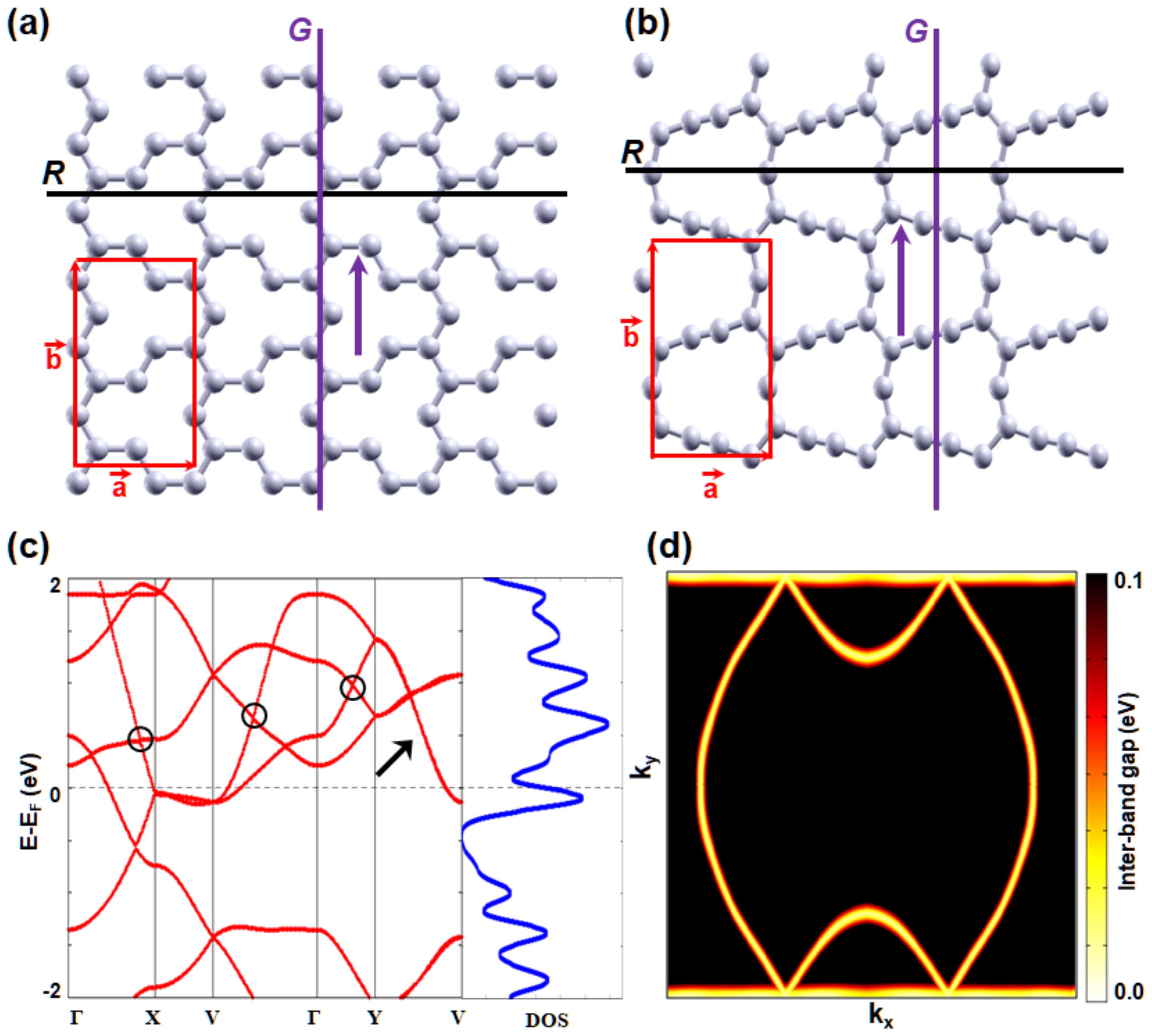}
    \caption{Lattice of C$_{10}$ (a) before and (b) after structural optimization. There are 10 C atoms in the unit cell defined by lattice vectors $\vec{a}$ and $\vec{b}$. The lattices have two symmetries: symmorphic reflection plane $R$ perpendicular to $\vec{b}$ and nonsymmorphic glide plane $G$ perpendicular to $\vec{a}$. (c) Band structure of C$_{10}$ along the high-symmetry line $\Gamma$-X-V-$\Gamma$-Y-V, with the Fermi level set at zero energy. The black arrow and circles indicate the symmetry-enforced nodal lines and the circles indicate accidental nodal lines, respectively. The corresponding density of states is shown to the right of the band structure. (d) Contour plot of the two kinds of nodal-lines by projecting them on the $k_x$-$k_y$ plane, with the color bar indicating the inter-band energy gap. The symmetry-enforced nodal line is at the $k_{y}=\pm\pi$ boundary of the Brillouin zone, while the accidental one is inside.}
    \label{fig:c10}
\end{figure}

 The lattices engineered from graphene are denoted by C$_N$, where $N$ is the number of sublattices in the rectangular unit cell.
We firstly consider a lattice that belongs to the wallpaper group $p2mg$\cite{Aroyo2016}, which has a glide plane going along ${\hat{\bf y}}$, and a reflection plane along ${\hat{\bf x}}$, as shown in Fig.~\ref{fig:c10} (a) for a C$_{10}$ configuration. In this case, we demonstrate that every two of the $N$ bands must stick together and form $N/2$ symmetry-enforced nodal lines at the BZ boundary $k_{y}=\pm\pi$. In contrast to the previous works about vacancy-engineered nodal lines that are based on analyzing how the pairwise degenerate eigenvalues of the nonsymmorphic symmetry operator constrain the band structure\cite{Liu21,Liu21_2}, in this work we present a new formalism based on a general feature of these nonsymmorphic vacancy configurations that enforces the nodal lines irrespective of the detail of the Hamiltonian. The general feature is that every two sublattices form a pair that map to each other under glide-plane operation, which we call a glide pair. Denoting the position of the unit cell to be $(x,y)$ and each sublattice to be $(A,B,C...)$, there are two kinds of glide pairs that we call type-I and type-II. The electron annihilation operators $c_{x,y}^{\rm sub}$ for a type-I/II glide pair $(A,B)/(C,D)$ transform under glide-plane operation as
\begin{eqnarray}
&&{\rm Type\;I}:\;c_{-x,-y}^{A}\rightarrow c_{x,-y}^{B},\;\;\;c_{-x,-y}^{B}\rightarrow c_{x,-y+1}^{A},
\nonumber \\
&&{\rm Type\;II}:\;c_{-x,-y}^{C}\rightarrow c_{x-1,-y}^{D},\;\;\;c_{-x,-y}^{D}\rightarrow c_{x-1,-y+1}^{C}.\;\;\;
\label{typeI_typeII_transformation}
\end{eqnarray}
The above transformations convey that the glide plane reflects the $x$-coordinate of a type-I pair, whereas in a type-II pair the glide plane reflects the $x$-coordinate and then translates to a neighboring unit cell along $x$. This translation is due to a shift between the center of the unit cell and the glide plane along the $x$-direction (c.f. Fig.~\ref{fig:c10} (a)). 
In both type-I and type-II pairs, the reflection of the $x$-coordinate is followed by a translation along $y$ which lands at the neighboring unit cell for the $B \rightarrow A$ and $D \rightarrow C$ transformations. The presence of the two orthogonal translations means that the glide plane is, simultaneously, a nonsymmorphic and an off-centered symmetry\cite{Liu21,Malard2018}.

With the transformations in Eq. (1) and arrangement of the basis functions according to the $N_{1}$ type-I and $N_{2}$ type-II glide pairs, the $N\times N$ glide-plane operator $G({\bf k})$ of the whole system is block diagonal, with $N_{1}+N_{2}=N/2$ blocks of $2\times 2$ matrices
\begin{eqnarray}
&&G({\bf k})=N_{1}\,g_{1}({\bf k})\oplus N_{2}\,g_{2}({\bf k})\Updownarrow_{k_{x}},
\nonumber \\
&&g_{1}({\bf k})=\left(\begin{array}{cc}
 & e^{-ik_{y}} \\
1 &
\end{array}\right),\;\;\;g_{2}({\bf k})=\left(\begin{array}{cc}
 & e^{-ik_{x}-ik_{y}} \\
e^{-ik_{x}} &
\end{array}\right),
\nonumber \\
\label{p2mg_Gk_form}
\end{eqnarray}
where $\Updownarrow_{k_{x}}$ takes $k_{x}$ to $-k_{x}$. The lattice described by the Bloch Hamiltonian $H({\bf k})$ has glide-plane symmetry if $\left[H({\bf k}),G({\bf k})\right]=0$. It then follows that there are $\left\{N_{1},N_{1},N_{2},N_{2}\right\}$-fold degenerate glide plane symmetry eigenvalues
$\left\{+e^{-ik_{y}/2},-e^{-ik_{y}/2},+e^{-ik_{x}-ik_{y}/2},-e^{-ik_{x}-ik_{y}/2}\right\}$. The simultaneous eigenstates of $G({\bf k})$ and $H({\bf k})$ satisfy
\begin{eqnarray}
&&G({\bf k})|\psi_{n1\pm}({\bf k})\rangle=\pm e^{-ik_{y}/2}\Updownarrow_{k_{x}}|\psi_{n1\pm}({\bf k})\rangle,
\nonumber \\
&&G({\bf k})|\psi_{n2\pm}({\bf k})\rangle=\pm e^{-ik_{x}-ik_{y}/2}\Updownarrow_{k_{x}}|\psi_{n2\pm}({\bf k})\rangle,
\nonumber \\
&&H({\bf k})|\psi_{nI\pm}({\bf k})\rangle=E_{nI\pm}(k_{x},k_{y})|\psi_{nI\pm}({\bf k})\rangle.\;\;\;\;\;\;
\label{single_glide_eigenvalues}
\end{eqnarray}
where $n=\left\{1,2...\right\}$ is the band index, $I=\left\{1,2\right\}$ stems from the two types of glide pairs. In Appendix \ref{apx:single_glide}, we elaborate that combining the appropriate transformation properties of the eigenstate with the fact that $k_{y}=0$ and $k_{y}=2\pi$ are the same point, we arrive at a condition for the eigenenergies
\begin{eqnarray}
&&E_{nI-}(k_{x},2\pi)=E_{nI+}(k_{x},0),
\nonumber \\
&&E_{nI+}(k_{x},2\pi)=E_{nI-}(k_{x},0),
\end{eqnarray}
meaning that at a given $\left\{k_{x},n,I\right\}$, the two bands with energy $E_{nI+}(k_{x},k_{y})$ and $E_{nI-}(k_{x},k_{y})$ swap and hence must cross each other somewhere in $0\leq k_{y}\leq 2\pi$. Applying the same argument to the BZ boundary $k_{y}=\pm\pi$ further dictates that the band crossing must occur at the BZ boundary
\begin{eqnarray}
E_{nI+}(k_{x},\pi)=E_{nI-}(k_{x},\pi).
\label{band_sticking_single_glide}
\end{eqnarray}
Thus every two bands with parameters $\left\{nI+\right\}$ and $\left\{nI-\right\}$ at any $k_{x}$ have to stick together at $k_{y}=\pi$, yielding $N/2$ nodal-lines there.

The validity of the above analysis is further supported by our first principles calculations for the band structure of various nonsymmorphic configurations in the $p2mg$ group. The band structures are obtained using the QUANTUM ESPRESSO package~\cite{Giannozzi2009}. The kinetic energies cutoff for wave function (ecutwfc) and for charge density (ecutrho) are set to 500 and 45 Ry, respectively. Perdew, Burke and Ernzerhof (PBE) form of the generalized gradient approximation (GGA) is adopted for the exchange-correlation energy~\cite{JPPerdew}. Numerical integrations in the BZ are evaluated with the Monkhorst-Pack mesh of $10\times10\times1$. All structures are relaxed until the total energy converges to within $10^{-4}$ eV during the self-consistent loop, with forces converged to 0.1 eV/nm, while employing the Methfessele-Paxton method with a smearing of 0.2 eV width. With the optimized geometry of the graphene structure and the corresponding self-consistent ground state computed with Quantum ESPRESSO, we use Wannier90 \cite{Mostofi2014} to map the ground-state wave functions onto a maximally localized Wannier function basis, and employ an adaptive $k$-mesh strategy to extract the matrices to build the real-space tight-binding model in the basis of the $s$, $p_x$, $p_y$ and $p_z$ orbitals of C atoms. We find that although the vacancy-engineered lattice distorts after the geometry optimization process as shown in Fig.~\ref{fig:c10}(b) for C$_{10}$, the relaxed lattice still belongs to the same $p2mg$ group and satisfies all the requirements in our argument.

Figure ~\ref{fig:c10} (c) displays the band structure of C$_{10}$ within a 2.0 eV window centered at the Fermi energy, obtained by density functional theory (DFT) calculation. Along the Y-V direction which corresponds to $k_{y}=\pi$, every two pairs of bands stick together to form a nodal line, as predicted. We note that, in the absence of spin-orbit coupling, all bands are completely spin-degenerate throughout the Brillouin zone (BZ). Hence, the nodal lines formed along the Y-V BZ edge are four-fold degenerate, with a two-fold degeneracy from spin and a two-fold degeneracy enforced by the glide-plane symmetry. In addition, from the DOS shown in Figure ~\ref{fig:c10} (c), we see that the DOS no longer vanishes linearly near the chemical potential as in pristine graphene\cite{Castro-2009}, but has a finite value due to the more complicated band structure. Figure~\ref{fig:c10}(d) shows the contour plot on the $k_{x}$-$k_{y}$ plane of the glide-plane-enforced nodal lines indicated by the arrow in Fig.~\ref{fig:c10}(c), as well as of accidental nodal lines indicated by the black circles. The latter arise from the accidental crossing of the splitting bands as they disperse from the symmetry enforced nodal lines. Accidental nodal lines have also been observed in holey graphene\cite{Chen-2018}, but the mechanism therein is unrelated to crystalline symmetry, and therefore the resulting nodal lines are unstable against SOC. In contrast, the accidental nodal lines shown in the interior of the BZ in Fig.~\ref{fig:c10}(d) arise from splitting the symmetry enforced ones, and hence are robust against SOC and any glide-plane symmetry-preserving perturbations.

\begin{figure}[ht]
\includegraphics[width=0.95\columnwidth]{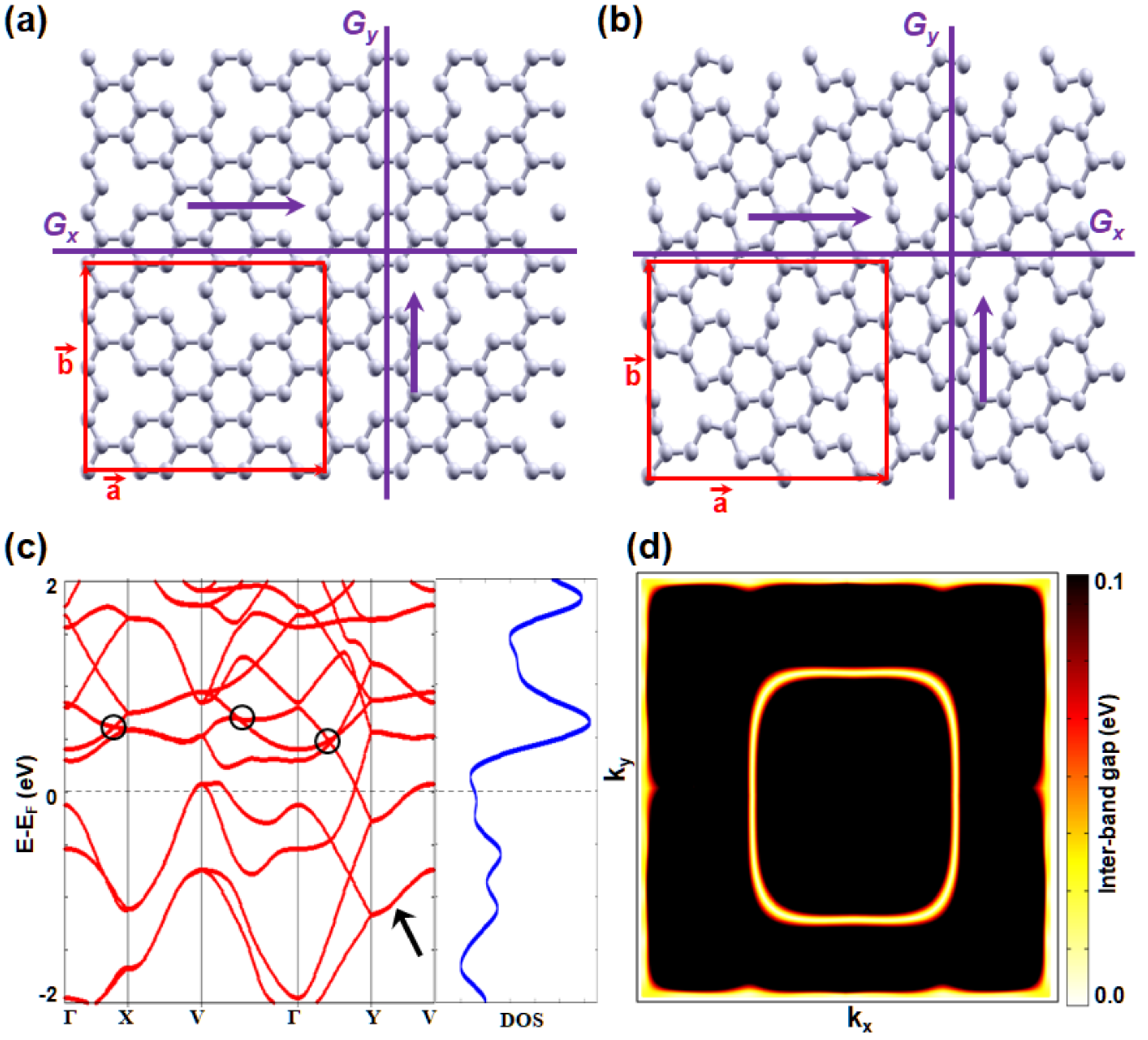}
    \caption{Lattice of C$_{44}$ (a) before and (b) after structural optimization. There are 44 C atoms in the unit cell defined by lattice vectors $\vec{a}$ and $\vec{b}$. The lattices have two orthogonal glide-planes $G_x$ and $G_y$. (c) Band structure of C$_{44}$ along the path $\Gamma$-X-V-$\Gamma$-Y-V, where the black arrows and circles indicate the symmetry-enforced and accidental nodal loops, respectively. The corresponding density of states is shown to the right of the band structure. (d) Contour plot of the two kinds of nodal loops by projecting them on the $k_x$-$k_y$ plane, with the color bar indicating the inter-band energy gap. The symmetry-enforced nodal loop surrounds the Brillouin zone boundary, while the accidental one is inside.}
    \label{fig:c44}
\end{figure}

\subsection{Vacancy-engineered graphene with two orthogonal glide planes \label{sec:two_glide_planes}} 

We proceed to discuss vacancy-engineered graphene that contain two orthogonal glide planes going along ${\hat{\bf x}}$ and ${\hat{\bf y}}$ directions, with the corresponding glide plane operators $G_{x}$ and $G_{y}$. Figure~\ref{fig:c44}(a) shows the lattice structure of C$_{44}$ which belongs to the wallpaper group $p2gg$. Even after the lattice relaxation, C$_{44}$ still hosts two orthogonal glide planes, as seen in Fig.~\ref{fig:c44}(b). Because the Hamiltonian commutes with both glide-plane operators, $\left[H({\bf k}),G_{x,y}({\bf k})\right]=0$, we are able to label the common eigenstates by the quantum numbers $\left\{n,\alpha,\beta\right\}$, where $n$ is the band index, $\alpha=I_{x}\pm$ labels the eigenvalues of $G_{x}$, and $\beta=I_{y}\pm$ labels the eigenvalues of $G_{y}$. Through
generalizing the argument for one glide plane to two orthogonal glide planes, as detailed in Appendix \ref{apx:two_glide}, we arrive at the condition
\begin{eqnarray}
&&E_{n\alpha I_{y}-}(k_{x},\pi)=E_{n\alpha I_{y}+}(k_{x},\pi),
\nonumber \\
&&E_{nI_{x}-\beta}(\pi,k_{y})=E_{nI_{x}+\beta}(\pi,k_{y}),\;\;\;\;
\end{eqnarray}
where $E_{n\alpha\beta}(k_{x},k_{y})$ are the eigenenergies. Thus every two bands are forced to stick together all around the BZ boundary, forming $N/2$ symmetry enforced nodal loops.

The above assertion is verified in the DFT band structure of C$_{44}$ in Fig.~\ref{fig:c44} (c) which clearly displays band crossings for every two pairs of spin-degenerate bands along the lines X-V and Y-V, yielding the outer four-fold degenerate nodal loop shown in the contour plot of Fig.~\ref{fig:c44} (d), and an enlarged DOS near the Fermi level. The band structure also exhibits accidental nodal loops inside the BZ, which are caused by crossings of bands as they split from the symmetry-enforced nodal-loops at the BZ boundary.


\subsection{Effect of intrinsic and Rashba SOC on nodal lines and nodal loops}

In this section, we use C$_{44}$ to elaborate that in the presence of Rashba SOC, even though the spin degeneracy of the band structure is lifted, the glide plane symmetry still forces every two spin-split bands to stick together at the BZ boundary, ensuring the existence of nodal lines and nodal loops, which are now two-fold degenerate. To demonstrate this effect, we consider the nearest-neighbor tight-binding model of graphene with Rashba SOC described by the Hamiltonian
\begin{eqnarray}
H&=&t\sum_{\langle ij\rangle,\sigma}c_{i\sigma}^{\dag}c_{j\sigma}
+i\lambda_{SOC}\sum_{\langle ij\rangle,\alpha,\beta}c_{i\alpha}^{\dag}\left({\boldsymbol \sigma}_{\alpha\beta}\times{\bf d}_{ij}\right)^{z}c_{j\beta}
\nonumber \\
&&+U\sum_{i\in vac,\sigma}c_{i,\sigma}^{\dag}c_{i\sigma}.\;\;\;
\label{Hamiltonian_graphene_Rashba}
\end{eqnarray}
Here $c_{i \sigma}$ is the electron annihilation operator of spin $ \sigma $ on the lattice site $ i$, $t$ is the hopping amplitude between nearest neighbor lattice sites $ \langle i j \rangle $, $\lambda_{SOC}$ is the coupling constant of Rashba SOC caused by breaking the inversion symmetry in the out-of-plane direction ${\hat{\bf z}}$, ${\boldsymbol \sigma}=(\sigma^{x},\sigma^{y},\sigma^{z})$ are the Pauli matrices, ${\bf d}_{ij}$ is the vector connecting site $i$ to site $j$. A very large on-site potential $U\sim 100t$ is applied on the vacancy sites $i\in vac$ to conveniently create the desired vacancy configuration. 



\begin{figure}[ht!]
\includegraphics[width=0.99\columnwidth]{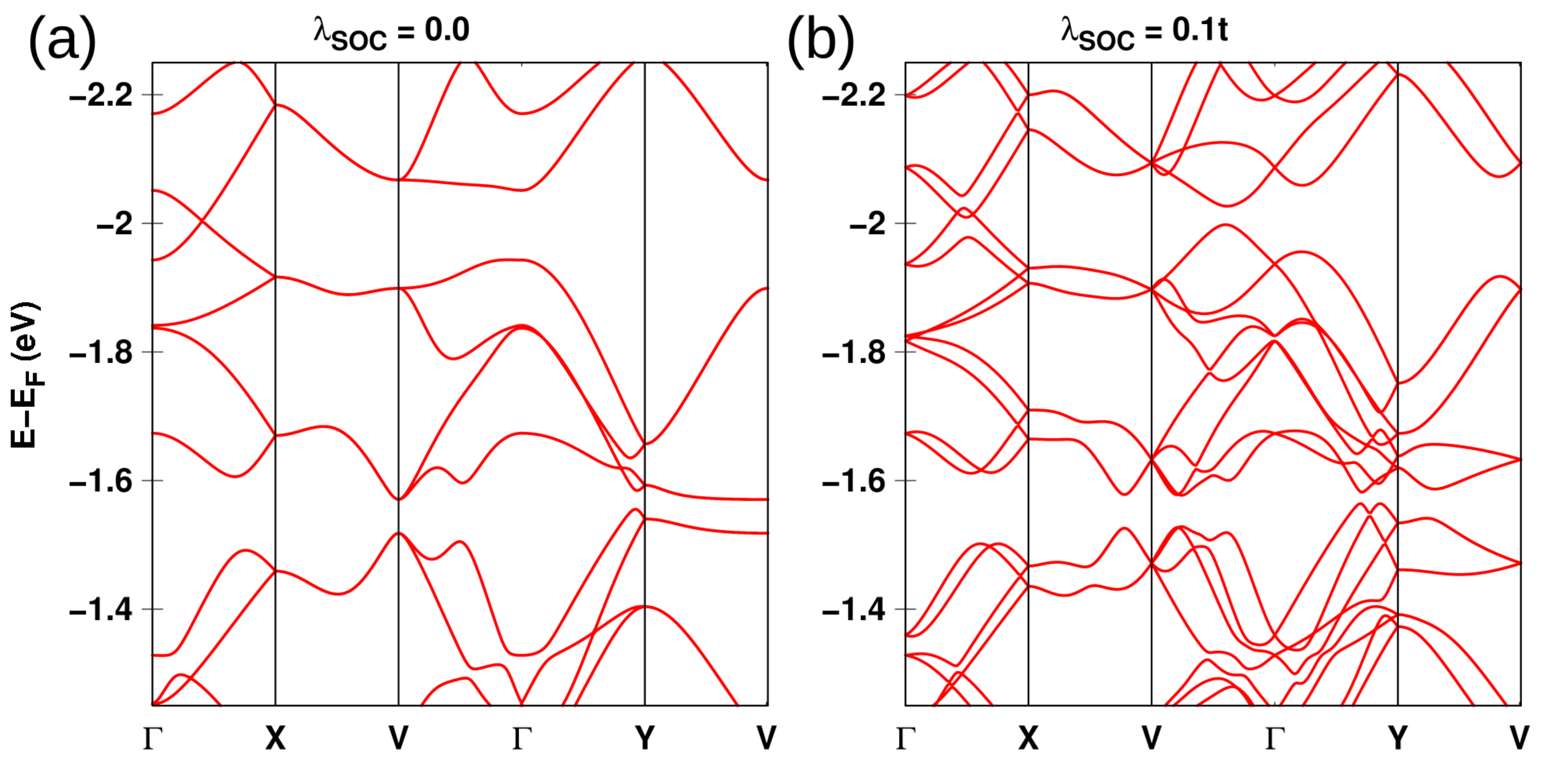}
    \caption{(a) The band structure of the C$_{44}$ configuration in Fig.~\ref{fig:c44} without Rashba SOC simulated by a nearest-neighbor tight-binding model. The nodal loops in the BZ boundary $X-V$ and $Y-V$ are four-fold degenerate. (b) The band structure of C$_{44}$ in the presence of Rashba SOC, which shows that despite the spin-splitting of the bands inside the BZ, every two bands still stick together at the BZ boundary to form nodal loops. The nodal loops are two-fold degenerate in this case due to the lifting of the spin degeneracy. }
    \label{fig:band_c44_glide_glide_rsoc}
\end{figure}

We will use this tight-binding model to examine the effect of Rashba SOC on the C$_{44}$ configuration in Fig.~\ref{fig:c44}, which has two orthogonal glide planes. In the pristine C$_{44}$ without Rashba SOC, the two orthogonal glide planes cause every two bands to stick together at the BZ boundary, and in addition there is spin degeneracy, so the nodal loops in Fig.~\ref{fig:c44} (a) are in fact four-fold degenerate. In contrast, Fig.~\ref{fig:band_c44_glide_glide_rsoc} (b) shows that at a finite Rashba SOC, the spin degeneracy is lifted everywhere inside the BZ as expected. Nevertheless, every two spin-split bands still merge together to form a two-fold degenerate nodal loop at the BZ boundary. In short, the Rashba SOC splits the spin degeneracy of the nodal loops and hence changes their degeneracy from four-fold to two-fold, but the glide-plane symmetry still ensures the existence of nodal loops at the BZ boundary.

\begin{figure}[ht]
\includegraphics[width=0.95\columnwidth]{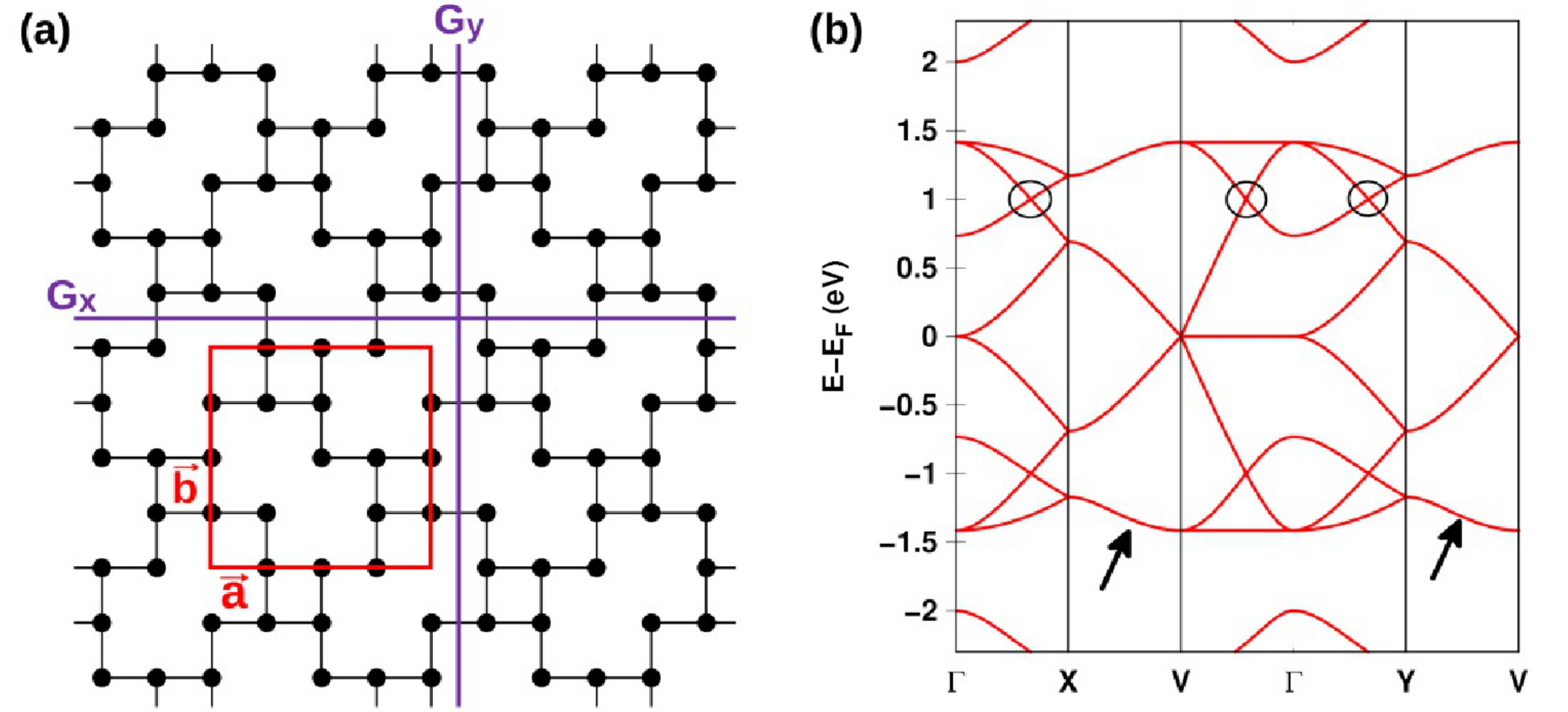}
\caption{(a) A vacancy configuration containing two orthogonal glide planes $G_{x}$ and $G_{y}$ engineered from a square lattice, and (b) its tight-binding band structure along high-symmetry lines. The black arrows and circles indicate the symmetry-enforced and accidental nodal loops, respectively.}
\label{fig:square_p4g_lattice_band_structure}
\end{figure}

\subsection{Vacancy-engineered NLSM from a square lattice}

In this section, we elaborate that our proposal is in fact a general principle not limited to graphene, but also applicable to other 2D materials with different lattice structures. Figure \ref{fig:square_p4g_lattice_band_structure} (a) shows a crystal structure vacancy-engineered from a square lattice, which belongs to wallpaper group $p4gm$ that contains
two orthogonal glide planes. The band structure obtained by tight-binding model with nearest-neighbor hopping $H=\sum_{\langle ij\rangle}t\,c_{i}^{\dag}c_{j}$ contains nodal-loops at BZ boundary just like C$_{44}$, as shown in Figure \ref{fig:square_p4g_lattice_band_structure} (b). This result indicates that our vacancy engineering principle can be generically applied to any 2D lattices regardless the structural and chemical details of the host system, and it is based solely on crystalline symmetries. 




\subsection{Experimental realization in nanoporous graphene}

Concerning the experimental realization of our proposal, a particularly promising route is the so-called bottom-up type of approach to nanoporous graphene. In this type of approach, one starts from small clusters of some precursor molecules, and choose a suitable chemical environment such that they self-assemble into lattice structures with periodic vacancies. This technique has been applied to grow graphene with periodic nanometer size pores\cite{Moreno18,Jacobse20}. In fact, the experimental vacancy configuration realized in Ref.~\onlinecite{Jacobse20}, called chevron-type nanoporous graphene (C-NPG), is nonsymmorphic. The lattice belongs to wallpaper group $p2gg$ that contains glide planes in two orthogonal crystalline directions, similar to the C$_{44}$ example discussed in Sec.~\ref{sec:two_glide_planes}. Therefore, the C-NPG should contain nodal loops surrounding the entire BZ edge according to our theory, as have also been confirmed by DFT calculations (see Fig.~3 E of Ref.~\onlinecite{Jacobse20}), although this feature has not been emphasized. However, this C-NPG contains a band gap $\sim$ eV at the Fermi level, and all the nodal loops form outside the band gap. Thus we anticipate that some experimental efforts is needed to search for other nanoporous configurations that contain nodal lines or loops crossing the Fermi level, such that the DOS may be enhanced instead of reduced.



\section{Conclusions} 

In summary, we elaborate that vacancy-engineered nonsymmorphic graphene exhibits band structures with multiple symmetry enforced nodal lines or nodal loops at the BZ boundary. This mechanism is based on the formation of glide pairs of the sublattices, which manifests regardless the original material is semimetallic, like graphene, or metallic, like a square lattice. In addition, accidental nodal lines and nodal loops can also occur inside the BZ. In fact, such a nonsymmorphic vacancy configurations have been realized experimentally in a nanoporous graphene\cite{Jacobse20}, and the existence of nodal loops in this configuration has been confirmed by DFT calculations, although it has not been emphasized. Our mechanism thus opens a new direction to explore vacancy-triggered NLSMs which can coexist with other material properties like Rashba SOC, and is even compatible with other types of vacancy-engineered band structures such as flat bands\cite{deSousa21_flatband}, hence may be exploited to fabricate novel NLSM-based electronic or spintronic devices. Moreover, shall the nodal lines cross the Fermi level, the finite DOS at the Fermi level is expected to dramatically alter thermal, electric and magnetic properties of the material compared to those in pristine graphene. We anticipate that this nonsymmorphic vacancy-engineering principle can be widely applied to change the band structure of a great variety of 2D materials, with the accompanying change of physical properties that awaits to be explored.


\acknowledgements

We thank exclusively D. Kochan for the discussion about SOC. W.C. is supported by the productivity in research fellowship from CNPq.

\appendix

\section{Nodal lines enforced by a single glide plane \label{apx:single_glide}}

We now give a detailed formalism for the nodal lines and nodal loops enforced by nonsymmorphic symmetry of the vacancy engineered lattices, starting from the nodal lines in the situation that the lattice contains only one glide plane. We will consider the spinless situation for simplicity, but the argument can be easily generalized to include spin. First we elaborate why the eigenvalues of $G({\bf k})$ in Eq.~(\ref{single_glide_eigenvalues}) contain the momentum-swapping operation $\Updownarrow_{k_{x}}$ using a simple example. Consider the minimal situation of $N=2$ sublattice as an example, in which $G({\bf k})=g(k_{y})\Updownarrow_{k_{x}}$, and we intend to diagonalize it to obtain the eigenvalues $\lambda_{\pm}({\bf k})$
\begin{eqnarray}
g({\bf k})|\phi_{\pm}({\bf k})\rangle=\lambda_{\pm}({\bf k})|\phi_{\pm}({\bf k})\rangle.
\end{eqnarray}
Denoting the eigenstate by
\begin{eqnarray}
|\phi_{\pm}({\bf k})\rangle=\left(\begin{array}{c}
u_{\pm}(k_{x},k_{y}) \\
v_{\pm}(k_{x},k_{y})
\end{array}\right),
\end{eqnarray}
the eigenvalue problem leads to
\begin{eqnarray}
&&g({\bf k})|\phi_{\pm}({\bf k})\rangle=\left(\begin{array}{cc}
 & e^{-ik_{y}} \\
1 &
\end{array}\right)\Updownarrow_{k_{x}}\left(\begin{array}{c}
u_{\pm}(k_{x},k_{y}) \\
v_{\pm}(k_{x},k_{y})
\end{array}\right)
\nonumber \\
&&=\left(\begin{array}{cc}
 & e^{-ik_{y}} \\
1 &
\end{array}\right)\left(\begin{array}{c}
u_{\pm}(-k_{x},k_{y}) \\
v_{\pm}(-k_{x},k_{y})
\end{array}\right)
\nonumber \\
&&=\lambda_{\pm}(k_{x},k_{y})\left(\begin{array}{c}
u_{\pm}(k_{x},k_{y}) \\
v_{\pm}(k_{x},k_{y})
\end{array}\right),
\end{eqnarray}
we see that this equation in general cannot be solved, because there needs not be a relation between $u_{\pm}(-k_{x},k_{y})$ and $u_{\pm}(k_{x},k_{y})$, or between $v_{\pm}(-k_{x},k_{y})$ and $v_{\pm}(k_{x},k_{y})$. Thus the correct way to diagonalize it is to maintain the $\Updownarrow_{k_{x}}$ in the eigenvalues
\begin{eqnarray}
g({\bf k})|\phi_{\pm}({\bf k})\rangle=\lambda_{\pm}({\bf k})\Updownarrow_{k_{x}}|\phi_{\pm}({\bf k})\rangle,
\end{eqnarray}
such that the diagonalization leads to
\begin{eqnarray}
&&g({\bf k})|\phi_{\pm}({\bf k})\rangle=\left(\begin{array}{cc}
 & e^{-ik_{y}} \\
1 &
\end{array}\right)\Updownarrow_{k_{x}}\left(\begin{array}{c}
u_{\pm}(k_{x},k_{y}) \\
v_{\pm}(k_{x},k_{y})
\end{array}\right)
\nonumber \\
&&=\left(\begin{array}{cc}
 & e^{-ik_{y}} \\
1 &
\end{array}\right)\left(\begin{array}{c}
u_{\pm}(-k_{x},k_{y}) \\
v_{\pm}(-k_{x},k_{y})
\end{array}\right)
\nonumber \\
&&=\lambda_{\pm}(k_{x},k_{y})\Updownarrow_{k_{x}}\left(\begin{array}{c}
u_{\pm}(k_{x},k_{y}) \\
v_{\pm}(k_{x},k_{y})
\end{array}\right)
\nonumber \\
&&=\lambda_{\pm}(k_{x},k_{y})\left(\begin{array}{c}
u_{\pm}(-k_{x},k_{y}) \\
v_{\pm}(-k_{x},k_{y})
\end{array}\right),
\end{eqnarray}
and hence one can solve for the coefficients $u_{\pm}(-k_{x},k_{y})$ and $v_{\pm}(-k_{x},k_{y})$ with eigenvalues $\lambda_{\pm}=\pm e^{-ik_{y}/2}$. This argument can be arbitrarily generalize to unit cells that contain more glide pairs.

We then consider the fact that, at a fixed $k_{x}$, the Hamiltonian at $k_{y}=0$ and $k_{y}=2\pi$ is the same, $H(k_{x},0)=H(k_{x},2\pi)$, and so is the glide-plane operator, $G(k_{x},0)=G(k_{x},2\pi)$. At a fixed $k_{x}$ and band index $n$, the symmetry eigenvalues at $k_{y}=0$ and $k_{y}=2\pi$ are,
\begin{eqnarray}
&&G(k_{x},0)|\psi_{n1\pm}(k_{x},0)\rangle=\pm\Updownarrow_{k_{x}}|\psi_{n1\pm}(k_{x},0)\rangle,
\nonumber \\
&&G(k_{x},2\pi)|\psi_{n1\pm}(k_{x},2\pi)\rangle=\mp\Updownarrow_{k_{x}}|\psi_{n1\pm}(k_{x},2\pi)\rangle,
\nonumber \\
&&G(k_{x},0)|\psi_{n2\pm}(k_{x},0)\rangle=\pm e^{-ik_{x}}\Updownarrow_{k_{x}}|\psi_{n2\pm}(k_{x},0)\rangle,
\nonumber \\
&&G(k_{x},2\pi)|\psi_{n2\pm}(k_{x},2\pi)\rangle=\mp e^{-ik_{x}}\Updownarrow_{k_{x}}|\psi_{n2\pm}(k_{x},2\pi)\rangle,
\nonumber \\
\end{eqnarray}
Combining this with $G(k_{x},0)=G(k_{x},2\pi)$ implies that one must be able to find a gauge in which
\begin{eqnarray}
&&|\psi_{nI+}(k_{x},0)\rangle=|\psi_{nI-}(k_{x},2\pi)\rangle,
\nonumber \\
&&|\psi_{nI-}(k_{x},0)\rangle=|\psi_{nI+}(k_{x},2\pi)\rangle.
\end{eqnarray}
It then follows that the eigenenergies satisfy
\begin{eqnarray}
&&H(k_{x},0)|\psi_{nI+}(k_{x},0)\rangle=E_{nI+}(k_{x},0)|\psi_{nI+}(k_{x},0)\rangle
\nonumber \\
&&=H(k_{x},2\pi)|\psi_{nI+}(k_{x},2\pi)\rangle
=E_{nI-}(k_{x},2\pi)|\psi_{nI-}(k_{x},2\pi)\rangle
\nonumber \\
&&=E_{nI-}(k_{x},2\pi)|\psi_{nI+}(k_{x},0)\rangle,
\label{Hkpsik_argument_k02pi}
\end{eqnarray}
since $H(k_{x},0)=H(k_{x},2\pi)$. This and a similar argument leads to
\begin{eqnarray}
&&E_{nI-}(k_{x},2\pi)=E_{nI+}(k_{x},0),
\nonumber \\
&&E_{nI+}(k_{x},2\pi)=E_{nI-}(k_{x},0).
\end{eqnarray}
Thus at given $k_{x}$, $n$, and $I$ the two bands $E_{nI+}(k_{x},k_{y})$ and $E_{nI-}(k_{x},k_{y})$ must cross each other somewhere in $0\leq k_{y}\leq 2\pi$.

We can apply the same argument to the BZ boundary $k_{y}=\pm\pi$, which has symmetry eigenvalues
\begin{eqnarray}
&&G(k_{x},\pi)|\psi_{n1\pm}(k_{x},\pi)\rangle=\mp i\Updownarrow_{k_{x}}|\psi_{n1\pm}(k_{x},\pi)\rangle,
\nonumber \\
&&G(k_{x},-\pi)|\psi_{n1\pm}(k_{x},-\pi)\rangle=\pm i\Updownarrow_{k_{x}}|\psi_{n1\pm}(k_{x},-\pi)\rangle,
\nonumber \\
\end{eqnarray}
and similarly for the eigenstate with index $I=2$. Because $G(k_{x},\pi)=G(k_{x},-\pi)$, there exists a gauge in which the eigenstates satisfy
\begin{eqnarray}
&&|\psi_{nI+}(k_{x},\pi)\rangle=|\psi_{nI-}(k_{x},-\pi)\rangle,
\nonumber \\
&&|\psi_{nI-}(k_{x},\pi)\rangle=|\psi_{nI+}(k_{x},-\pi)\rangle.
\end{eqnarray}
Using $H(k_{x},\pi)=H(k_{x},-\pi)$, the same procedure in Eq.~(\ref{Hkpsik_argument_k02pi}) leads to
\begin{eqnarray}
E_{nI+}(k_{x},\pi)=E_{nI-}(k_{x},-\pi)=E_{nI-}(k_{x},\pi),
\end{eqnarray}
where in the last equality we have used the fact that $k_{y}=\pi$ and $k_{y}=-\pi$ are the same point on the boundary of a rectangular BZ, thus completing the proof to Eq.~(\ref{band_sticking_single_glide}).

\begin{figure}[ht]
\includegraphics[width=0.99\columnwidth]{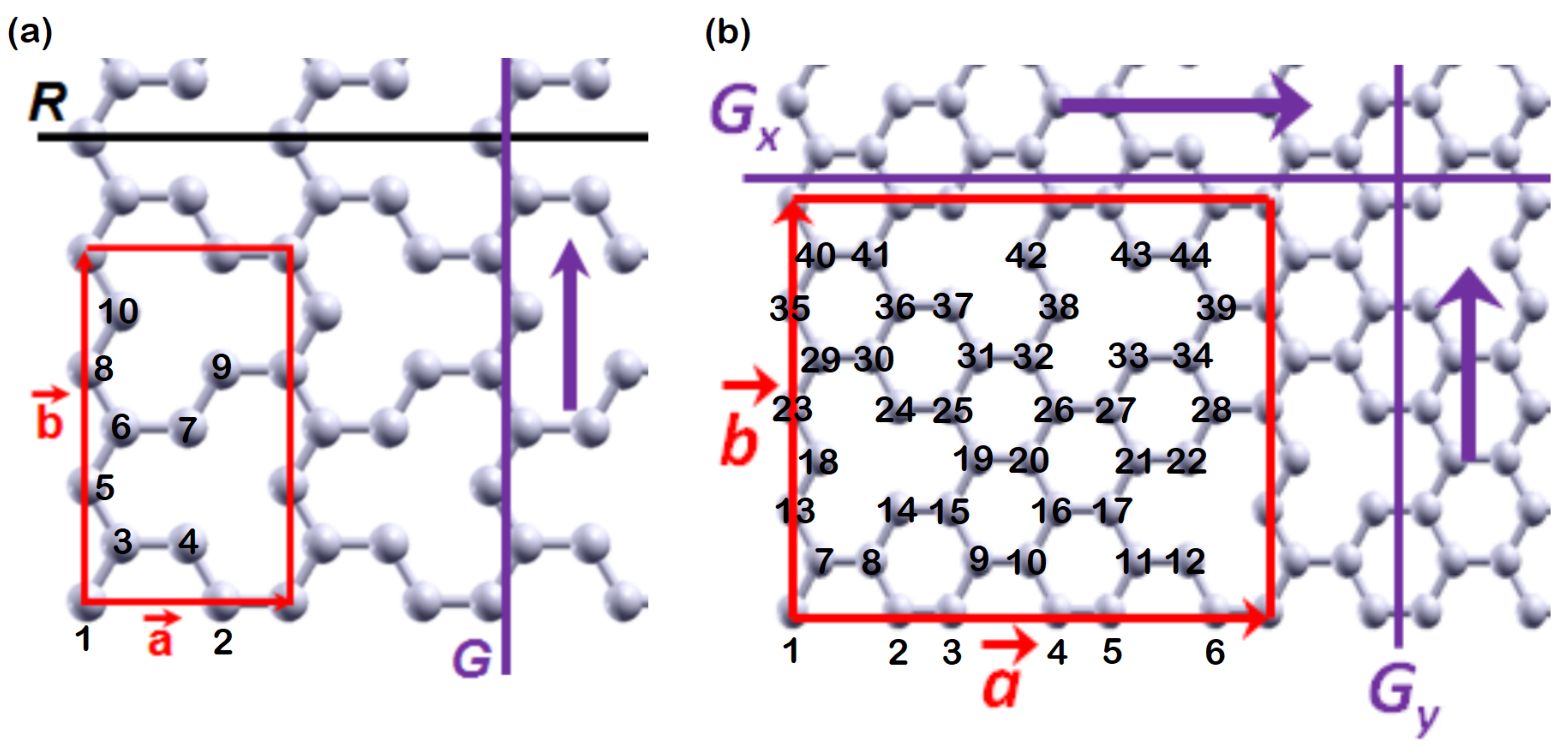}
    \caption{The numbering of sublattices for (a) the C$_{10}$ example in Fig.~\ref{fig:c10} and (b) the C$_{44}$ example in Fig.~\ref{fig:c44}. }
    \label{fig:C10_C44_sublattices}
\end{figure}

To be more concrete about the notion of glide pairs, using the numbering of sublattices and the glide vector in Fig.~\ref{fig:C10_C44_sublattices} (a), the glide pairs defined with respect to the glide plane $G$ for the C$_{10}$ configuration in Fig.~\ref{fig:c10} are
\begin{eqnarray}
{\rm Type-I}:\;&&(1,6),\;\;\;(3,8),\;\;\;(5,10),
\nonumber \\
{\rm Type-II}:\;&&(2,7),\;\;\;(4,9).
\label{C10_typeI_typeII}
\end{eqnarray}
For each of the 3 type-I pairs, mapping from left to right under $G$ remains in the same unit cell, but mapping from right to left under $G$ moves to the next unit cell in $+{\hat{\bf y}}$ direction; For each of the 2 type-II pairs, mapping from left to right moves to the next unit cell in $-{\hat{\bf x}}$ direction while from right to left moves to the next unit cell along $-{\hat{\bf x}}+{\hat{\bf y}}$.
As result, the glide plane operator is that in Eq.~(\ref{p2mg_Gk_form}) with $N_{1}=3$ and $N_{2}=2$, and so follows the discussion in this section.

\section{Nodal-loops enforced by two orthogonal glide planes \label{apx:two_glide}}

We proceed to discuss vacancy engineered graphene that belong to the wallpaper groups that have two orthogonal glide planes denoted by $G_{x}$ and $G_{y}$. In these wallpaper groups, a specific sublattice $A$ is mapped to another one $B$ under $G_{x}$, but it is mapped to a different one $C$ under $G_{y}$. In other words, the glide pair arrangements are different for $G_{x}$ and $G_{y}$. Thus if we arrange the basis according to the glide pairs of $G_{y}$, then $G_{y}$ will take the block-diagonal form of Eq.~(\ref{p2mg_Gk_form}),
but $G_{x}$ will not be block-diagonal in this basis because it has a different glide pair assignment. Nevertheless, $G_{x}$ will have $\left\{N_{1x},N_{1x},N_{2x},N_{2x}\right\}$ degenerate eigenvalues according to the numbers of type I and type II glide pairs defined for this glide plane, and $G_{y}$ will have $\left\{N_{1y},N_{1y},N_{2y},N_{2y}\right\}$ degenerate eigenvalues regardless how the basis is arranged. Because the Hamiltonian commutes with both of them, $\left[H({\bf k}),G_{x}({\bf k})\right]=0$ and $\left[H({\bf k}),G_{y}({\bf k})\right]=0$, one must be able to label the eigenstates by the quantum numbers $\left\{n,\alpha,\beta\right\}$, where $n$ is the band index, $\alpha=I_{x}\pm$ labels the eigenvalues of $G_{x}$, and $\beta=I_{y}\pm$ labels the eigenvalues of $G_{y}$. The eigenstates satisfy
\begin{eqnarray}
&&G_{x}({\bf k})|\psi_{n1\pm\beta}({\bf k})\rangle=\pm e^{-ik_{x}/2}\Updownarrow_{k_{y}}|\psi_{n1\pm\beta}({\bf k})\rangle,
\nonumber \\
&&G_{x}({\bf k})|\psi_{n2\pm\beta}({\bf k})\rangle=\pm e^{-ik_{x}/2-ik_{y}}\Updownarrow_{k_{y}}|\psi_{n2\pm\beta}({\bf k})\rangle,
\nonumber \\
&&G_{y}({\bf k})|\psi_{n\alpha 1\pm}({\bf k})\rangle=\pm e^{-ik_{y}/2}\Updownarrow_{k_{x}}|\psi_{n\alpha 1\pm}({\bf k})\rangle,
\nonumber \\
&&G_{y}({\bf k})|\psi_{n\alpha 2\pm}({\bf k})\rangle=\pm e^{-ik_{x}-ik_{y}/2}\Updownarrow_{k_{x}}|\psi_{n\alpha 2\pm}({\bf k})\rangle,
\nonumber \\
&&H({\bf k})|\psi_{n\alpha\beta}({\bf k})\rangle=E_{n\alpha\beta}({\bf k})|\psi_{n\alpha\beta}({\bf k})\rangle.
\end{eqnarray}
Following the same argument for the $p2mg$ group in the previous section, we obtain
\begin{eqnarray}
&&E_{n\alpha I_{y}\mp}(k_{x},2\pi)=E_{n\alpha I_{y}\pm}(k_{x},0),
\nonumber \\
&&E_{nI_{x}\mp\beta}(2\pi,k_{y})=E_{nI_{x}\pm\beta}(0,k_{y}),
\end{eqnarray}
implying a band crossing in the range $0\leq k_{x}\leq 2\pi$ at any fixed $k_{y}$, and another band crossing in the range $0\leq k_{y}\leq 2\pi$ at any fixed $k_{x}$. The argument applied to the BZ boundary also leads to
\begin{eqnarray}
&&E_{n\alpha I_{y}-}(k_{x},\pi)=E_{n\alpha I_{y}+}(k_{x},-\pi)=E_{n\alpha I_{y}+}(k_{x},\pi),
\nonumber \\
&&E_{nI_{x}-\beta}(\pi,k_{y})=E_{nI_{x}+\beta}(-\pi,k_{y})=E_{nI_{x}+\beta}(\pi,k_{y}).\;\;\;\;
\end{eqnarray}
Thus every two bands are forced to stick together at the BZ boundary, forming $N/2$ symmetry enforced nodal loops surrounding the BZ boundary.

For the C$_{44}$ example in Fig.~\ref{fig:c44}, using the numbering of sublattices in Fig.~\ref{fig:C10_C44_sublattices} (b), the glide pairs defined with respect to the glide plane $G_{x}$ are
\begin{eqnarray}
{\rm Type-I}:\;&&(1,10),\;\;\;(2,11),\;\;\;(3,12),\;\;\;(7,4),
\nonumber \\
&&(8,5),\;\;\;(9,6),
\nonumber \\
{\rm Type-II}:\;&&(13,42),\;\;\;(14,43),\;\;\;(15,44),\;\;\;(40,16),
\nonumber \\
&&(41,17),\;\;\;(18,38),\;\;\;(19,39),\;\;\;(35,20),
\nonumber \\
&&(36,21),\;\;\;(37,22),\;\;\;(23,32),\;\;\;(24,33),
\nonumber \\
&&(25,34),\;\;\;(29,26),\;\;\;(30,27),\;\;\;(31,28).
\nonumber \\
\end{eqnarray}
For the 6 type-I pair, mapping from left to right under $G_{x}$ remains in the same unit cell, but from right to left moves to the next unit cell in ${\hat{\bf x}}$ direction; For the 14 type-II pairs, mapping from left to right moves to the next unit cell in the $-{\hat{\bf y}}$ direction, whereas mapping from right to left moves to the next unit cell in the ${\hat{\bf x}}-{\hat{\bf y}}$ direction. As a result, the glide plane operator is that defined in Eq.~(\ref{p2mg_Gk_form}) with $N_{1x}=6$ and $N_{2x}=14$ and swapping $\left\{k_{x},k_{y}\right\}\rightarrow\left\{k_{y},k_{x}\right\}$. On the other hand, the glide pair assignment is different for the glide plane $G_{y}$, which are
\begin{eqnarray}
{\rm Type-I}:\;&&(1,26),\;\;\;(2,25),\;\;\;(3,24),\;\;\;(4,23),
\nonumber \\
&&(7,32),\;\;\;(8,31),\;\;\;(9,30),\;\;\;(10,29),
\nonumber \\
&&(13,38),\;\;\;(14,37),\;\;\;(15,36),\;\;\;(16,35),
\nonumber \\
&&(18,42),\;\;\;(19,41),\;\;\;(20,40),
\nonumber \\
{\rm Type-II}:\;&&(5,28),\;\;\;(6,27),\;\;\;(11,34),\;\;\;(12,33),
\nonumber \\
&&(17,39),\;\;\;(21,44),\;\;\;(22,43),
\end{eqnarray}
where the mapping follows that described after Eq.~(\ref{C10_typeI_typeII}), yielding the glide plane operator $G_{y}$ given by Eq.~(\ref{p2mg_Gk_form}) with $N_{1y}=15$ and $N_{2y}=7$, and so follows the discussion in this section.

\end{document}